# Coherent multidimensional spectroscopy of dilute gas-phase nanosystems

Lukas Bruder[1]*, Ulrich Bangert[1], Marcel Binz[1], Daniel Uhl[1], Romain Vexiau[2], Nadia Bouloufa-Maafa[2], Olivier Dulieu[2] and Frank Stienkemeier[1]

[1]Institute of Physics, University of Freiburg, 79104 Freiburg, Germany
[2]Laboratoire Aimé Cotton, CNRS, Université Paris-Sud, ENS Cachan, Université Paris-Saclay, 91405 Orsay Cedex, France

*Correspondence to: lukas.bruder@physik.uni-freiburg.de

June 21, 2018

**Two-dimensional electronic spectroscopy (2DES) is one of the most powerful spectroscopic techniques, capable of attaining a nearly complete picture of a quantum system including its couplings, quantum coherence properties and its real-time dynamics[1-3]. While successfully applied to a variety of condensed phase samples[4-10], high precision experiments on isolated quantum systems in the gas phase have been so far precluded by insufficient sensitivity. However, such experiments are essential for a precise understanding of fundamental mechanisms and to avoid misinterpretations, e.g. as for the nature of quantum coherences in energy transport[11,12]. Here, we solve this issue by extending 2DES to isolated nanosystems in the gas phase prepared by helium nanodroplet isolation in a molecular beam-type experiment. This approach uniquely provides high flexibility in synthesizing tailored, quantum state-selected model systems of single and many-body properties[13,14]. For demonstration, we deduce a precise and conclusive picture of the ultrafast coherent dynamics in isolated high-spin $Rb_2$ molecules and present for the first time a dynamics study of the system-bath interaction between a single molecule (here $Rb_3$) and a superfluid helium environment. The results demonstrate the unique capacity to elucidate prototypical interactions and dynamics in tailored quantum systems and bridges the gap to experiments in ultracold quantum science.**

The development of multidimensional spectroscopy in the optical regime has considerably advanced our understanding of microscopic processes, as it has improved the time-frequency resolution to an unprecedented level[2]. This technique maps the system's third order nonlinear response onto 2D frequency-correlation maps which provide invaluable information over one-dimensional methods. Key advantages are the high spectro-temporal resolution, the direct disclosure of couplings and the differentiation of homogeneous and inhomogeneous broadening mechanisms[1-3]. As such, 2DES has provided new insights in topics as broad as energy relaxation pathways and quantum coherence in photosynthetic systems[4-7], many-body correlations and exciton dissociation in semiconductor materials[8,9] and reaction pathways in photophysical/-chemical reactions[10].

Despite the success of 2DES, the vast complexity of investigated condensed phase systems makes precise analysis and modelling extremely difficult. This has led to some ambiguities in interpretations, most prominently the observation of long-lived quantum coherences in biological systems[11,12]. 2DES studies of single, isolated systems in the gas phase, e.g. in molecular beams, would strongly reduce the complexity and a first step in this direction has been recently reported[15]. However, the preparation of molecular complexes/aggregates in the gas phase is technically very restricted. Consequently, the aspect of inter-particle couplings/dynamics and environmental effects, being the essence of most microscopic processes, cannot be probed in such studies. Here, we resolve these issues by introducing a new concept combining 2DES with helium nanodroplet isolation (HENDI). This approach uniquely enables 2DES studies of isolated tailor-made model systems ideally suited to study intra- as well as intermolecular properties/dynamics and the influence of a controlled environment.

HENDI has been established as a unique technique for spectroscopic studies of atoms, molecules and their complexes, that are isolated in a superfluid helium matrix[13]. Spectroscopy of pigment molecules[16], up to larger biomolecules[17] and exotic species[14] has been demonstrated with a resolution often clearly exceeding other methods[13]. Translational and internal degrees of freedom of embedded species are efficiently cooled to sub Kelvin (370 mK) temperatures[13]. Heterogeneous complexes and molecular aggregates are readily synthesized directly in the droplets[18-22]. The rare-gas environment provides a prototypical perturbation which is much simpler to model than the influence of molecular solvent networks, and by co-doping with individual atoms/molecules (microsolvation), environmental parameters can be tuned in a controlled manner[21].

We demonstrate the basic principle and advantages of combining 2DES with HENDI in a model study of high-spin $Rb_2$ and $Rb_3$ molecules isolated on the surface of helium nanodroplets. In terms of sensitivity, requirements are similar for larger molecules/clusters, but for demonstrating high resolution at dilute conditions, simple molecular structures are preferable. Furthermore, alkali-metal molecules are of general high interest as they play an important role in ultracold physics and chemistry[23] and provide an ideal test bench for *ab initio* quantum chemistry methods[24-26]. The $Rb_2$ and $Rb_3$ molecules are prepared in their ground states (lowest vibrational level) directly on the surface of the helium droplets (Fig.1A, details in Materials section). Thereby, the droplets serve as a cold, inert substrate assisting the molecule formation and natural selection of the van der Waals-bound high-spin configurations $Rb_2$ $a^3\Sigma_u^+$ and $Rb_3$ $1^4A'_2$. These states are otherwise difficult to access due to their low binding energy $E_B$ ($Rb_2$ $E_B=235$ cm$^{-1}$,[25] $Rb_3$ $E_B=939$ cm$^{-1}$ [26]), hence, exemplifying the ability of HENDI for tailored molecular synthesis.

A major challenge in gas-phase experiments are the low target densities, making the application of advanced nonlinear spectroscopy methods extremely demanding. Especially, in HENDI experiments, densities are only $\leq 10^7$ cm$^{-3}$ (corresponds to optical density (OD) ~$10^{-11}$). The routinely employed experimental implementation of 2DES based on non-collinear four-wave-mixing is not suitable for such low molecular densities. To achieve the required high sensitivity, we instead use a collinear geometry, rapid phase modulation combined with efficient lock-in detection[27] and photoionization for detection (Fig. 1B, details in Material section).

In this scheme, four phase-modulated laser pulses induce a fourth-order nonlinear population in the sample (Fig. S3 in SI) which is mapped onto the photoionization yield. The weak nonlinear signal contributions are extracted from detected photoelectron or mass-resolved photoions based on their individual phase modulation signatures. This procedure is similar to phase-cycling, however, is performed here at an update rate of 200 kHz, and is therefore more sensitive than most pulse shaper-based setups. Likewise, photoionization is of advantage as it ensures higher collection efficiencies than photon detection, allows for selective probing through different ionization channels and provides additional information, e.g. about dark states, ion-mass information or ion/electron angular distributions.

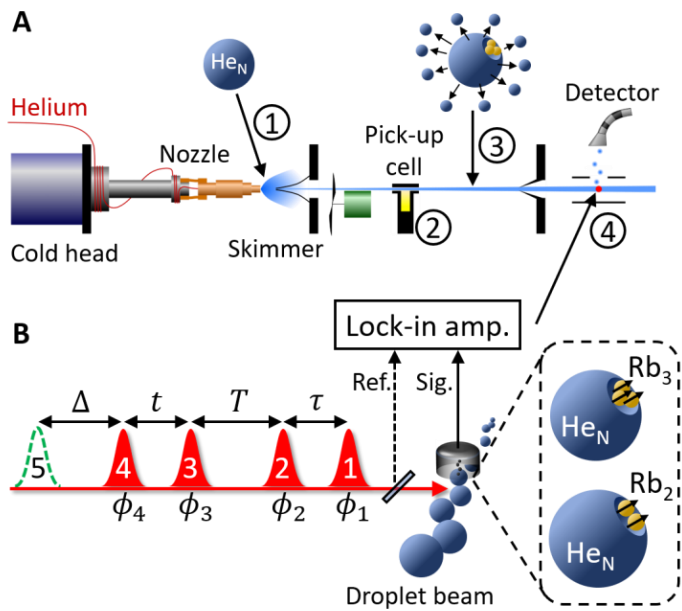

**Figure 1. Experimental scheme.** (**A**) Molecular beam-type vacuum apparatus for helium nanodroplet beam generation upon adiabatic expansion of helium (1), followed by doping with Rb atoms in a pick-up cell (2) and evaporative cooling of the formed Rb molecules (3). The isolated, cold $Rb_2$ and $Rb_3$ molecules intersect with the laser beam and photoelectrons/-ions are detected (4). (**B**) Four phase-modulated laser pulses with delays $\tau$, $T$ and $t$, excite and ionize the prepared molecules in the droplet beam. The phase $\phi_i$ of each pulse is individually modulated at kHz-frequencies, leading to a modulation-beat of the photoelectron/-ion yield. A lock-in amplifier is used for demodulation and isolation of the nonlinear 2D signal components. For a more selective ionization, a fifth pulse delayed by $\Delta$ is optionally applied.

Photoelectron-2D spectra of the isolated molecular species are shown in Fig. 2A, B and an ion-detected 2D spectrum is shown in C. These 2D maps directly correlate the pump excitation ($\omega_\tau$-axis, comparable to absorption spectrum) with the system response probed as a function of the evolution time $T$ ($\omega_t$-axis, comparable to emission spectrum). Thereby, the encoded phase information enables clear discrimination of signal contributions: ground state bleach (GSB)/stimulated emission (SE) both positive and excited state absorption (ESA) negative amplitude (details in SI). Furthermore, the peak magnitude strongly depends on the ionization scheme. This enables selective enhancement/discrimination of individual features (demonstrated in Fig. 2C), which, in contrast to previous 2DES experiments, provides us an additional means to disentangle the system response.

Considering the extremely low molecular densities in the experiment, the acquired 2D spectra reveal very high quality. The sharp, well-separated spectral features allow us unambiguous identification of spectral components and correlations among those. Absorption and emission profiles of the data (pump/probe projections, Fig. S4 in SI) are in excellent agreement with *ab initio* calculations[24,25] and high-resolution steady-state laser spectroscopy[18,19], which confirms the high fidelity of our method and facilitates clear assignment of all spectral features. At $\omega_\tau = 14125$ cm$^{-1}$ we observe the $Rb_3$ $1^4A_2' \rightarrow 1^4A_{1,2}''$ quartet transition and correlated negative ESA peaks, as well as a transient positive cross peak ($T=0$ fs). Around $\omega_\tau = 13500$ cm$^{-1}$, we observe the $Rb_2$ $a^3\Sigma_u^+ \rightarrow (1)^3\Pi_g$ triplet resonance with clearly resolved spin-orbit (SO) components $0_g^\pm, 2_g$ of the excited state along the $\omega_\tau$-axis[28]. Correlated to this resonance, we identify two ESA and one cross peak at off-diagonal positions (labeled ESA$_1$, ESA$_2$ and CP). The high resolution allows us furthermore to identify a distinct Stokes shift (red shift of diagonal peaks along $\omega_t$-axis, see also Fig. 3A). This asymmetry in absorption and emission is characteristic for molecules initially prepared in a single vibrational ground state[29], confirming the preparation of cold molecules in our experiment.

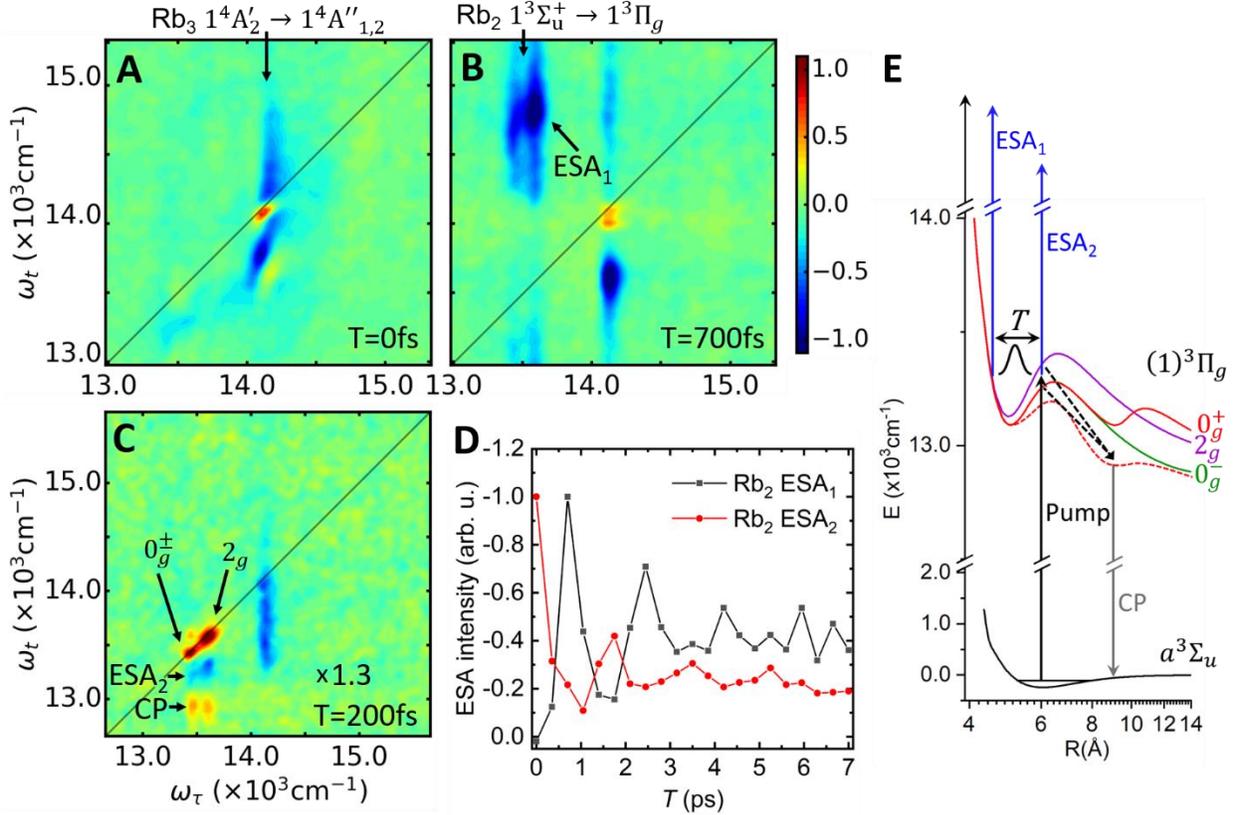

**Figure 2. Rb$_2$ and Rb$_3$ 2DES results.** (**A, B**) Photoelectron-2D correlation spectra of isolated Rb$_2$ and Rb$_3$ molecules for evolution times $T$=0 fs and 700 fs, respectively. Labels indicate the assigned transitions. (**C**) Selective enhancement of Rb$_2$ features using a wavelength-optimized fifth pulse combined with photoion detection. (**D**) Coherent oscillation of Rb$_2$ ESA peaks as a function of $T$. (**E**) Rb$_2$ PECs[24] and concluded photo dynamics. Transitions are labeled in accordance to (B, C). A droplet-induced blue shift of 115 cm$^{-1}$ is applied to the $(1)^3\Pi_g$ states and the influence of the helium perturbation on the $0_g^+$ state is indicated as dashed curve[30].

As an intriguing aspect, we observe a persistent coherent oscillation of the Rb$_2$ ESA peaks as a function of the evolution time $T$ (Fig. 2D). While in past 2DES studies, off-diagonal beat signals have been controversially discussed being either of electronic or vibronic nature[11,12], we can unambiguously identify our signal as a vibrational wave packet prepared by the pump step in the $(1)^3\Pi_g, \Omega = 0_g^\pm, 2_g$ excited states and probed via the ESA transitions (Fig. 2E). Franck-Condon (FC) calculations reveal a wave packet excitation around $v$=7 with an average level spacing of $\approx$ 20 cm$^{-1}$ (Fig. S5 in SI) in very good agreement with the observed oscillation period of $\approx$1550 fs. As an additional information, the initial phase of the ESA$_1$ and ESA$_2$ oscillation indicates two separate FC windows at the inner/outer turning points of the $(1)^3\Pi_g$ potential energy curves (PECs, Fig. 2E) which completes the picture of the wave packet dynamics.

Interestingly, the CP feature oscillates in phase with the ESA$_2$ peak (not shown) and reveals a considerably red-shifted emission ($\omega_t$ =12950 cm$^{-1}$), which is identical for all SO components, thus indicating an ultrafast relaxation at the outer turning point of the $0_g^\pm, 2_g$ states to a common lower-lying state. Vibrational relaxation within these states would not reproduce the observed red shift and no other electronic states are in spectral vicinity. Likewise, previous experiments assigned the red-shifted emission to gas-phase Rb$_2$ molecules[18] which clearly contradicts our time-resolved data. Therefore, we explain the CP feature by a relaxation/tunneling into the outer potential well of the $0_g^+$ state[24], catalyzed by the perturbation of the helium environment. The electron distribution of this state differs from the other bound $(1)^3\Pi_g$ states which might lead to a slightly smaller Rb$_2$-He$_N$ interaction (Fig. 2E) and thus would explain the red-shifted emission[30]. No clear oscillation can be resolved in the Rb$_2$ diagonal peaks within our signal quality. This might be due to mainly GSB pathway contributions which propagate on the electronic ground state.

The Rb$_2$ study has demonstrated our ability to experimentally disentangle ultrafast molecular dynamics with high precision and in excellent agreement with theory. We focus now on the dynamics in the Rb$_3$ system, where the spectrally isolated quartet resonance $1^4A_2' \to 1^4A_{1,2}''$ ($\omega_\tau$ =14125 cm$^{-1}$) serves us as a probe for system-bath interactions. Here, a second peak emerges from the Rb$_3$ diagonal peak (Fig. 3A), showing a dynamic red shift in the emission frequency, which converges to a constant shift of $\Delta\omega$=150±19 cm$^{-1}$ within 2.5 ps (see also Fig. S6 in SI). The ESA peak below the diagonal shows a similar red shift, however, compromised by an overlapping positive cross peak.

We interpret our data with a dynamic energy shift induced by the helium matrix (Fig. 3B). Upon excitation, the molecule's electron distribution expands, causing a repulsion and rearrangement of the surrounding helium density (bubble effect[31]), which leads to an increasing molecule-droplet distance $z$ and thus to a time-dependent reduction of the matrix-induced energy shift. This dynamic is reflected in the SE/ESA pathways as they evolve on the excited state of the Rb$_3$-He$_N$ potential, whereas the GSB pathway evolves on the ground state and shows no dynamic (Fig. S3 in SI). Note that the system-bath interaction is here repulsive (leading to a blue shift) whereas in the condensed phase, interactions are typically attractive and lead to red shifts.

At $T\approx$ 2.5 ps, the interaction potential curvature has reached a low gradient, explaining the constant SE/ESA peak position and the reduced line broadening along the emission axis of the SE, whereas at the same time, the GSB contribution exhibits an approximately symmetric peak shape. For $T \geq$ 50-100 ps, the SE/ESA peaks vanish in contrast to the GSB contribution (not shown) indicating the desorption and accompanied dissociation of the metastable Rb$_3$ quartet molecule. Similar desorption time scales have been deduced for Rb atoms and Rb$_2$ molecules[32,33]. Furthermore, the observed matrix shift of $\Delta\omega$=150±19 cm$^{-1}$ for Rb$_3$ is along the line of the shifts for Rb atoms ($\Delta\omega$=12 cm$^{-1}$)[34] and Rb$_2$ molecules ($\Delta\omega$=115 cm$^{-1}$, see SI). Respectively, we deduce for the gas-phase asymptote of the $1^4A_2' \to 1^4A_{1,2}''$ transition a value of 13938±16 cm$^{-1}$, which was so far unknown at this precision. Likewise, we deduce for the first time temporal information about the rearrangement dynamics of the superfluid helium surface, indicating a time scale < 2.5 ps.

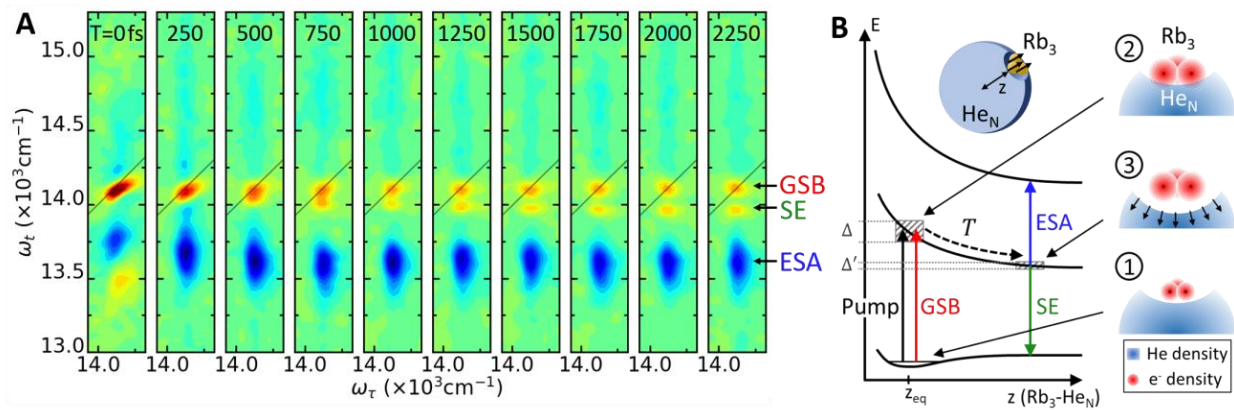

**Figure 3. Matrix-induced dynamic energy shift of the Rb₃ resonance.** (**A**) Time evolution of spectral features correlated to the Rb₃ $1^4A'_2 \to 1^4A''_{1,2}$ excitation. For a compact presentation, only a cutout of acquired 2D spectra are shown. (**B**) Schematic of the Rb₃-He$_N$ potentials explaining the matrix-induced dynamic shift. Step 1-3 sketch the expansion of the Rb₃ electron orbital upon excitation, followed by a repulsion of the helium density. In SE and ESA pathways, the system evolves on the excited state during *T*, leading to a dynamic red shift and peak narrowing on the probe axis, which is not present in the GSB pathway. $\Delta, \Delta'$ indicate the change in line broadening along the $\omega_t$-axis.

Due to strong neighboring cross and ESA peaks flanking the Rb₂ diagonal peak (Fig. 2C) and presumably dominating GSB contributions, we were not able to resolve this matrix effect there. We further note, that an excited wave packet passing through a conical intersection may yield a very similar dynamic picture as the one observed. However, this model contradicts available *ab initio* PECs[26].

In this work, we have presented the first 2DES study of isolated molecular nanosystems in the gas phase. Our results confirm that we can synthesize different species in a controlled manner, prepare them in well-defined initial states and retrieve high-resolution 2D data with fine spectral details, hard to observe in the condensed phase. By introducing photoionization to 2D spectroscopy, we achieve unprecedented sensitivity and demonstrate selective probing as an additional tool to disentangle the system response. This unique combination of methods enabled a precise analysis of our data even without the requirement of in-depth simulations, as shown for the photodynamics in Rb₂ attached to the droplet surface and the Rb₃-helium droplet interaction dynamics. Both molecules are of great interest in ultracold science and we add here new information about their femtosecond dynamics.

In the future, the possibility to study isolated tailor-made model systems combined with the developed highly selective 2D spectroscopy method will be extremely valuable in answering fundamental questions in primary photophysical and -chemical processes. Thereby, the role of the environment can be studied by co-doping with solvent molecules, using other rare-gas matrices or investigating isolated molecules in a seeded molecular beam, all readily implementable in our apparatus. The demonstrated high sensitivity will also open 2D spectroscopy studies of other fundamental gas-phase systems, e.g. mass-selected cluster beams, ultracold quantum gases or ion crystals.

## Methods

### Sample preparation:

Helium nanodroplet beam generation is described in detail elsewhere[35]. An extended sketch of our vacuum apparatus is shown in the SI (Fig. S1) along with the formation mechanism of Rb molecules. $^4$He gas (purity grade 6.0) is continuously expanded through a nozzle (5 µm diameter) cooled to 17 K with a stagnation pressure of 50 bar, leading to a mean droplet size of 7000 helium atoms. Upon condensation and evaporation, the droplets cool down to 370 mK[13] and undergo a phase transition into the superfluid phase. The droplet beam passes through a temperature-controlled pick-up cell (1 cm length, T= 377 K) containing a low-density Rb vapor ($3.9 \times 10^{-4}$ mbar). Alkali-metal atoms do not immerse into the droplets[34]. Pick-up of multiple atoms thus leads to molecule/cluster formation on the droplet surface. Thereby, the released binding energy is effectively dissipated upon evaporation of helium atoms, assisting the formation of the weakly-bound lowest high-spin electronic states of $Rb_2$ and $Rb_3$ molecules accompanied with cooling to their vibrational ground level. The considerably larger energy release of the low-spin electronic ground states of Rb molecules leads to an enhanced evaporation or droplet destruction, causing a predominance of the high-spin Rb molecules in the experiment[36].

### Optical setup and data acquisition:

The 2DES optical setup is based on the phase modulation (PM) technique developed by Marcus and coworkers[27]. Details are found in the SI. Briefly, four phase-modulated collinear pulses are focused (f=300 mm) into the interaction region of the detector to induce the nonlinear signals (Fig. 1B). Ionization is either performed with a separate fifth pulse (delayed by $\Delta \approx 2$ ns) or by absorbing additional photons from pulse 4. Independent wavelength tuning of pulse 5 allows for selective amplification/discrimination of specific spectral features. Photoelectrons or mass-resolved photoions are detected with a channeltron detector. For each evolution time $T$, the coherence times $\tau$ and $t$ are scanned and Fourier transformed afterwards to yield 2D frequency-correlation spectra of which the real part is shown. To isolate the weak nonlinear signal components, the phase $\phi_i$ of each pulse is individually modulated (Fig. 1B) at radio frequencies $\Omega_i$, leading to characteristic modulation signatures for the rephasing ($\Omega_{RP} = \Omega_{43} - \Omega_{21} = 3$ kHz) and non-rephasing ($\Omega_{NRP} = \Omega_{43} + \Omega_{21} = 13$ kHz) third order signals which are separated upon lock-in detection. Thereby, amplitude and phase information are retrieved through heterodyned detection by referencing the lock-in amplifier to a suitable reference. The excitation scheme along with double-sided Feynman diagrams and phase signatures of pathway contributions is shown in the SI (Fig. S3).

## Data availability

The data that support the findings of this study are available from the corresponding author upon reasonable request.

## Acknowledgements

We thank A.W. Hauser for fruitful discussions about the $Rb_3$ molecule and W.E. Ernst for providing us the $Rb_2$ and $Rb_3$ laser spectroscopy data. Funding by the European Research Council (ERC) within the Advanced Grant "COCONIS" (694965), by the Deutsche Forschungsgemeinschaft (DFG) IRTG CoCo (2079) and the use of the computing center MésoLUM of LUMAT research federation (FR LUMAT 2764) is acknowledged.

## Author contributions

F.S. and L.B. conceived the experiment. U.B., M.B., D.U. and L.B. implemented the experiment. U.B. and M.B. performed the measurements. O.D., R.V., N.B-M. and L.B. performed the $Rb_2$ calculations. L.B. wrote the manuscript with input from all other authors.

## Competing interests

The authors declare no competing financial interests.